\documentstyle [11pt] {article}
\date{}
\begin{document}

\title{MICROWAVE BACKGROUND ANISOTROPIES, LARGE-SCALE STRUCTURE AND
COSMOLOGICAL PARAMETERS}

\author{ A. KASHLINSKY\\
Code 685, Laboratory for Astronomy and Solar Physics\\Goddard Space
Flight Center, Greenbelt, MD 20771, USA}

\maketitle
\noindent
{\bf to appear in proceedings of Yamada conference XXXVII on ``Evolution of
the Universe..." (Tokyo)}\\
\noindent

\section*{Abstract}
We review how the various large-scale data constrain cosmological parameters
and, consequently, theories for the origin of large-scale structure in the
Universe. We discuss the form of the power spectrum implied by the correlation
data of galaxies and argue by comparing the velocity field implied by the
distribution of light with the observed velocity flows that the bias parameter,
$b$,
is likely to be constant in the linear regime. This then allows one to estimate
the density parameter, $\Omega$,
and $b$ directly from the \underline{data} on $\xi(r)$ and
the velocity fields. We show that it is consistent with low values of
$\Omega^{0.6}/b$. We discuss the ways to normalise the optical data at $z\sim0$
directly to the COBE (or other microwave background) data. The data on high-$z$
\underline{galaxies} allows one to further constrain the shape of the
\underline{primordial} power spectrum at scales which are non-linear today
($< 8h^{-1}$Mpc) and we discuss the consistency of the data with inflationary
models normalised to the large-scale structure observations.

\section{Introduction}
The purpose of this review is to discuss the constraints the current
astronomical data place on the values of cosmological parameters, such
as $\Omega$, the \underline{primordial} power spectrum $P(k)$ etc. Indeed,
the last few years brought wealth of observational data in optical, radio and
and microwave bands that allow now to constrain these parameters more tightly
and thus to gain further insight into the early Universe physics, particularly
in light of the inflationary picture.
We first discuss the implications for the inflationary scenario of
the (realistic) possibility that the Universe may turn out to be open.
We point out that the Grischuk-Zeldovich (1978) effect combined
with the COBE observations of the quadrupole
anisotropy of the microwave background radiation (MBR) would then preclude the
possibility that the Universe's homogeneity was produced by inflationary
expansion during its early evolution. Next, in Section 3 we discuss the
constraints on the \underline{primordial} power spectrum of the
\underline{light} distribution on scales $10-100
h^{-1}$Mpc from the optical data at $z\sim0$. Section 4
deals with comparing the (mass) power spectrum deduced from the peculiar
velocities data and that of the light distribution. We point out that the
two are consistent with each other and therefore it is indeed plausible to
assume that they are proportional to each other, i.e. the bias factor $b$
is constant. We further note that the comparison of the two leads to low
values of $\Omega^{0.6}/b$. In Section 5 we discuss the normalisation
of the power spectrum to the COBE results and whether the latter necessarily
imply flat Universe and/or Harrison-Zeldovich spectrum of the primordial
density fluctuations. Section 6 discusses the constraints on $P(k)$ on small
scales, which are non-linear today, from the data on high redshifts
\underline{galaxies} and the Uson et al (1992) object. We conclude in Section
7.
\section{$\Omega$ and inflation}
The value of the density parameter, or more precisely the curvature radius
of the Universe, is certainly the most critical test of inflation. Some
data seems consistent with the flat Universe (Kellerman 1993 and these
proceedings; Yahil 1993,these proceedings). But I think it is fair to say
that most observational data seems to point to low values of $\Omega$. In
particular the age ($t_0$) measurements indicate that $H_0t_0 > \frac{2}{3}$
(e.g. Lee 1992 and these proceedings) and dynamics of the Local Group strongly
prefers $\Omega \sim 0.1-0.2$ (Peebles 1989 and Tully 1993, these proceedings;
see also Kashlinsky 1992a and discussion in Sec.4).

What would it mean if the \underline{data} were to prove that the Universe is
old and open? There were some suggestions in the past attempting to accomodate
the open Universe with the
finely-tuned inflation (e.g. Ellis 1988; Steinhardt 1990), the idea being
that the Universe has undergone only the minimal numbers of $e$-foldings
(of order $\sim$65) necessary to make it homogeneous on scales not much
greater than the present horison, $R_{hor}
\sim 6000h^{-1}$Mpc. This would then make the curvature radius $\sim R_{hor}$
leading to $\Omega$ as low as $\sim 0.1-0.3$, but would also imply that we
just happened to be born at the time when the horison did not yet grow large
enough to encompass more than the inflation-blown homogeneous bubble.

The problem with such finely-tuned inflationary models is that they would
violate the COBE measurements of the quadrupole anisotropy
as discussed by Frieman, Kashlinsky and Tkachev (1993) (see also Turner 1991).
In that case the Universe would be inhomogoneous on scales greater than the
horison. If the amplitude of the superhorison harmonics of lengthscale $l$
is $h$, it would cause, via the Grischuk-Zeldovich (1978) effect, the
quadrupole microwave background anisotropy of the amplitude
$Q \simeq h(R_{hor}/l)^2$. Now both $l$ and $\Omega$ are, in inflationary
scenarios, functions of the number of $e$-foldings, $N$, while the amplitude
of $Q$ produced by the superhorison inhomogenieties cannot exceed the COBE
found value of $Q_{COBE} \simeq 5\times 10^{-6}$.
Thus one can express $l$ in terms of $\Omega$ and then
rewrite the constraint imposed by the Grischuk-Zeldovich
effect \underline{and} the COBE data
directly as a constraint on the present value
of the density parameter (Frieman, Kashlinsky and Tkachev 1993):
\begin{equation}
1-\Omega \leq \frac{Q_{COBE}}{h} \simeq O(10^{-6})
\end{equation}
The above means that if the Universe went through an inflationary phase
during its evolution (so that $h \gg 1$ on scales where inflationary smoothing
was inefficient), the COBE \underline{observations} require it to have the
density parameter within a factor $\sim 10^{-6}$ of unity.
Conversely, if it turns out
that observations prove that the Universe is open this would mean
that the homogeniety of the Universe has not originated by the inflationary
expansion.
\section{Structure at $z\sim 0$ and the primordial power spectrum}
The distribution of galaxies and galaxy systems (light) is now measured
fairly accurately
on scales up to $\sim 100h^{-1}$Mpc from various and independent datasets. All
the datasets give results which are in good agreement with each other and, if
the light distribution is at least proportional to that of mass and all find
substantially more (light) power on large scales than simple inflationary
models (e.g. cold-dark-matter - CDM) would predict. As
was mentioned, the data discussed
in this section measure the distribution of light and in order to get
information about the mass power spectrum one has to make assumptions about
the interdependence between the light and mass distributions.

We discuss below in chronological order the most accurate datasets
for determining the power spectrum of light on large scales:

1) Cluster-cluster correlation function was measured for Abell (Bahcall and
Soneira 1983) and Zwicky (Postman et al 1986) clusters. The data showed
that the correlation function of light remains positive on scales up to
$\simeq 100h^{-1}$Mpc and is roughly $\xi(r) \propto r^{-2}$ thus implying
the power spectrum of light, $P(k) \propto k^n$, of $n\simeq -1$ (Kaiser 1984;
Kashlinsky 1987,1991a). Furthermore, the data showed a systematic increase of
the correlation amplitude with the cluster richness (or mass). This as
suggested
by Kashlinsky (1987,1991a) can be explained within the gravitational clustering
model of cluster formation and the dependence of the increase on cluster masses
requires $n \simeq -1$ and is inconsistent with the standard CDM model
(Kashlinsky 1991a). There were some suggestions that the cluster-cluster
correlation is strongly biased on large scales by projection effects
(Sutherland
1988), but there is some evidence to the contrary (e.g. Szalay et al 1989).
Furthermore, since the power
implied by it is consistent with other and later findings discussed below,
there is good chance it reflects a true distribution of light, if not of the
mass itself.

2) The two-point galaxy correlation function of galaxies has been measured
in the APM survey (Maddox et al 1991). The measurements were done in the
$b$-band and covered approximately one square steradian of $\sim$2.5 million
galaxies. This is probably the most accurate measurement of the projected
angular correlation function $w(\Theta)$ with the systematic error being
less than $2\times 10^{-3}$. The results show significantly more power than
the standard CDM model would predict and imply $n<0$ on scales $<100h^{-1}$Mpc.

3) Picard (1991) has compiled the POSS catalogue in the $r$-band
of aprroximately 400,000
galaxies in projection and determined $w(\Theta)$ which is roughly
consistent with the APM survey. Similarly, the COSMOS machine results give
$w(\Theta)$ in good agreement with the APM data and the cluster-cluster
correlations (Collins et al 1992).

4) Redshift surveys allow one to map the galaxy distribution in the 3-D space,
but here one has to correct for the distortions induced by the peculiar
velocity flows (Kaiser 1988). Vogeley et al (1992) have measured the power
spectrum of the light distribution from the CfA redshift survey and
find results consistent with the ones mentioned above. Fisher et al
(1993) find similar results from the catalogue of $\sim$5,000 IRAS galaxies.

5) QDOT analysis of counts in cells (Saunders et al 1991) for
$\sim$2,000 IRAS galaxies gives results consistent with the above.

Thus the consistency of the above independent studies of different objects,
in different wavebands and at different depths indicates a good probability
of the
reality of these results. Consequently it may make little sense to discuss
cosmological models (e.g. CDM) in the context of one set of mearuments only
(e.g. COBE). One has to compare theories with \underline{all} observational
data, i.e. to normalise them simultaneously to the large-scale data, that
map the distribution of matter at the present epoch ($z\sim0$) and MBR
observations which map the mass distribution at the last scattering surface
(in the absence of reheating $z\sim 1100$). We devote the rest of this review
to discussing how such normalisation can be done (Juszkiewicz et al 1987;
Gorski 1991; Kashlinsky, 1991b;1992a,b,c) and what information it carries.

So what is the power spectrum (of light) implied by the large-scale structure
data? Below, we use mainly the APM data, but as discussed above the other
datasets would give consistent results. For small angles $\Theta$ the two-point
correlation function $w(\Theta)$ decreases with $\Theta$ as $\Theta^{-\gamma}$
with $\gamma\simeq 0.7$. This implies a power spectrum $P(k) \propto
k^{\gamma-2}$ for sufficiently large $k$ (small scales). On larger angular
scales $w(\Theta)$ falls off rapidly and its signal becomes lost in the
systematic noise ($\simeq 2\times 10^{-3}$). The fall-off may imply the
following things: 1) the power spectrum goes into the white-noise regime
($n=0$) leading to $\xi(r)\simeq0$; 2) $P(k)$ goes into the Harrison-Zeldovich
regime, $n=1$, where the correlation function is negative and falls off rapidly
with $r$: $\xi(r) \propto -r^{-4}$; 3) the power spectrum has the power index
$n$ even greater than the Harrison-Zeldovich value of 1 in which case the
correlation function decreases even more rapidly with $r$: $\xi(r) \propto
-r^{n+3}$ and the signal gets lost in the noise. Thus the only conclusion one
can make from the fall-off in $w(\Theta)$ at large $\Theta$ is that
these scales correspond to the power index $n\geq0$ at sufficiently small $k$.
Various forms for the fit were proposed by Kashlinsky (1991a,b) and Peacock
(1991).
Below we will adopt the form from Kashlinsky (1992c) which is particularly
simple to use for quick estimates and which gives essentially the same results
as the former two:
\begin{equation}
P(k) = \frac{2 \pi^2 \xi(r_8)}{k_0^3 \Phi_\xi(k_0 r_8; n)} \times
\left\{ \begin{array}{ll}
		(\frac{k}{k_0})^n, & k<k_0\\
		(\frac{k}{k_0})^{\gamma-2}, & k>k_0
	\end{array}
\right.
\end{equation}
The above has been normalised to to the observed value, $\xi(r_8)=
(r_8/r_*)^{-\gamma-1}$ at $r_8=8h^{-1}$Mpc and $r_*=5.5h^{-1}$Mpc and
$\Phi(y,n)$$\equiv$$\int_0^1x^{2+n}j_0(xy)dx$$+$$\int_1^\infty x^{\gamma}
j_0(xy)dx$.
The value of the transition scale, $k_0^{-1}$, must be determined from matching
eq.(2) to the APM data on $w(\Theta)$. Note, that because of the rapid fall-off
in $\xi(r)$ for $n\geq0$, the value of $k_0^{-1}$ is rather insensitive to the
precise value of the free parameter $n$ as long as it remains positive on
large scales. Kashlinsky (1992c) has analysed the APM
data in the narrow magnitude slices, each slice $\Delta m\simeq 0.5$ wide,
thereby reducing effects due to evolution of galaxies lying at different
depths and finds $k_0^{-1}=50h^{-1}$Mpc from matching (2) to the APM data.
Peacock's (1991) form for $P(k)$ would give similar results when applied to
the non-scaled APM data in the narrow magnitude bins.
\section{Normalising to velocity field: does light trace mass?}
How is the power spectrum of the light distribution discussed in the previous
section related to the power spectrum of mass, which if one accepts the
inflationary prejudices is uniquely specified by the early Universe physics?
Obviously within the framework of inflationary models, the transition scale
to the Harrison-Zeldovich regime is not a free parameter and is approximately
equal to the horison scale at the epoch of the matter-radiation equallity.
It is thus expected that for such models $k_0^{-1}$ should be $\simeq  13
(\Omega h)^{-1}h^{-1}$Mpc, so that the correlation function of mass
for the standard CDM model ($\Omega=1, h=1/2$) has zero crossing at $\simeq
30h^{-1}$Mpc. This is considerably smaller than the value of $k_0^{-1}$
indicated by the data.

The answer to the above
question comes from comparing the power spectrum implied
by the peculiar velocity data (which map the \underline{mass} distribution)
with that of light discussed in the previous section. This has been done by
Kashlinsky (1992a), who proposed a method to relate the velocity field directly
to the correlation function, thereby eliminating the power spectrum entirely
from discussion. He then computed the peculiar velocity field implied directly
by the APM data on $w(\Theta)$ (assuming that the latter is proportional to
the mass power spectrum) and compared the results with the Great Attractor
peculiar field. The comparison is plotted on Fig.2 of Kashlinsky (1992a), which
shows that the peculiar velocity field (its amplitude \underline{and} coherence
length) due to the APM data is in good agreement with the Great Attractor
data. This therefore suggests that it is indeed reasonable to assume that at
least in the linear regime (scales $r>8h^{-1}$Mpc) the bias factor, $b$, can
(and must?) be assumed to be constant with scale.

Furthermore, using the methods developed in Kashlinsky (1992a) one can
determine
the density parameter $\Omega$ and the bias factor $b$ by comparing the
peculiar velocity data with the velocity field predicted by the APM. For
simplicity, we reformulate this method in terms of the three-dimensional
correlation function, $\xi(r)$. The ``dot" peculiar velocity correlation
function is defined as $\nu(r) =<\!\mbox{\boldmath$v$}(\mbox{\boldmath$x$})
\cdot \mbox{\boldmath$v$}(\mbox{\boldmath$x$}+\mbox{\boldmath$r$})\!>$, and,
if the light and mass power spectra are proportional to each
other, the power spectrum can be eliminated from discussion
and one can relate $\nu(r)$ directly
to $\xi(r)$ (see Kashlinsky 1992a for details). The relation between the two is
given by the second order differential equation:
\begin{equation}
\nabla^2 \nu(r) = -\frac{\Omega^{1.2}}{b^2} H_0^2 \xi(r)
\end{equation}
As discussed in the previous section, the APM data suggest that the zero
crossing of $\xi(r)$ occurs on large scale,
$\simeq 2.5k_0^{-1}\simeq 130h^{-1}$Mpc. Thus, as follows
from eq.(2), on scales, on which the velocity
data is most reliably determined ($\leq 60-70h^{-1}$Mpc) $\xi(r)$ is to a good
approximation given by $\xi(r) \simeq (r/r_*)^{-\gamma-1}$. Using this
approximation for $\xi(r)$ (rooted in \underline{observations}), one can
solve eq.(3)
analytically using as one boundary condition the fact that $\nu(0)$ must be
finite. Defining $V_* \equiv H_0r_* = 550$km/sec, the solution is given by:
\begin{equation}
\nu(r) = \nu(0) - \frac{\Omega^{1.2}}{b^2} \frac{V_*^2}{(1-\gamma)(2-\gamma)}
(\frac{r}{r_*})^{1-\gamma}
\end{equation}
The observed value of
the dot velocity correlation function at zero is given by e.g.
pairwise velocities, cluster velocity dispersions etc and is thought to be
$\sqrt{\nu(0)}\simeq 500-700$km/sec (e.g. Peebles 1987 and references cited
therein). The lower end of that range would be in better agreement with the
dipole MBR anisotropy (local) motion of $\simeq 630$km/sec.
The data on $\nu(r)$ were determined on scales $\leq 60h^{-1}$Mpc by
Bertschinger et al (1990) and we use their values at two linear
scales, 40 and 60$h^{-1}$Mpc, to determine from eq.(4) the values of $\nu(0)$
implied by the data for various values of $\Omega^{0.6}/b$. The results are
shown in the Table 1 below for $\gamma=0.7$.\vspace{0.3cm}\\
Table 1.
\begin{tabular}
{c | c  c  c }  & $\Omega^{0.6}/b$ & {$r=40h^{-1}$Mpc} & {$r=60h^{-1}$Mpc} \\
 $\sqrt{\nu(r)}$ &  & 388 km/sec & 330 km/sec \\
\hline
 & 1 & 1250 km/sec & 1300 km/sec \\
$\sqrt{\nu(0)}$ & 0.5 & 708 km/sec & 709 km/sec \\
 & 0.3 & 526 km/sec & 500 km/sec\\
\hline
\end{tabular}
\vspace{0.3cm}\\
The second row shows the values for the data for $\sqrt{\nu(r)}$ at 40 and
60$h^{-1}$Mpc used in computing $\sqrt{\nu(0)}$ shown in Table 1 against the
values of $\Omega^{0.6}/b$. One can see from the Table that the values of
$\Omega^{0.6}/b$ preferred by the data on $\xi(r), \nu(r)$ and $\nu(0)$ are:
\begin{equation}
(0.2-0.3) < \frac{\Omega^{0.6}}{b} < (0.4-0.5)
\end{equation}
This analytical estimate
is consistent with our earlier results (Kashlinsky 1992a) and the
results presented by Brent Tully at this meeting, but disagrees with the POTENT
determination of these parameters (Yahil 1993, these proceedings).

\section{Normalising to COBE}

In order to determine/constrain $P(k)$ one must use the large scale data
discussed in the previous sections in conjunction with the COBE observations
on the MBR correlation function $C(\theta)$. However, such determination
can at present be made reliably only on scales below the curvature radius,
$R_{curv}=cH_0^{-1}/\sqrt{1-\Omega}$. Since the smallest scales subtended by
COBE ($\sim 10^o$) exceed the curvature radius for values of $\Omega$ even as
high as $\Omega \simeq 0.2-0.3$ (and the quadrupole scale always exceeds
$R_{curv}$ in open Universe) this analysis can be done only for the case of
flat Universe. We therefore, concentrate in this section on discussing how well
the standard inflationary scenario fits \underline{both} the APM/peculiar
velocity data which restrict $P(k)$ on scales $< 100h^{-1}$Mpc and the COBE
data
which subtend scales $>$$600h^{-1}$Mpc if $\Omega$=1.
In its conventional form, the inflationary
scenario makes two predictions: 1) the Universe must be flat ($\Omega$$=$$1$)
to a very high accuracy, and 2) the \underline{initial} spectrum of the
primordial density fluctuations  must have the Harrison-Zeldovich
form, $P(k)$$\propto$$k$. In the standard model fluctuations do not grow
on sub-horison scales during the radiation dominated era; on larger scales
the growth would be self-similar. This leads to a uniquie shape
of the transfer function
accounting for the modification of the power spectrum, such that
the Harrison-Zeldovich
shape must be preserved on sufficiently large scales. (Constraints on the
transfer function or $P(k)$ on small scales are discussed in the next section).

The first-year COBE data give the values of
$\sqrt{C_{10^o}}$, the signal convolved with the $10^o$ FWHM beam and the
quadrupole anisotropy $Q$. For $P(k)$ given by (2) with the Harrison-Zeldovich
spectrum ($n=1$) and normalised to the APM
data via $k_0$ we obtain within the uncertainty of the bias factor:
\begin{equation}
\sqrt{C_{10^o}(0)} \simeq
\frac{2.6\times 10^{-5}}{b} \frac{k_0^{-1}}{50h^{-1}Mpc}
\; \; ; \; \;
Q \simeq \frac{1.2\times 10^{-5}}{b} \frac{k_0^{-1}}{50h^{-1}Mpc}
\end{equation}
One can see that it is possible to fit the COBE results if the bias factor is
sufficiently large, $b$$>$2. Indeed, the intrinsic unceratinty in the APM data
would probably restrict $k_0$ to lie in the range 40$h^{-1}$Mpc$<$$k_0^{-1}$$
<$60$h^{-1}$Mpc with $k_0^{-1}$=50$h^{-1}$Mpc being the best fit.
COBE data give $\sqrt{C_{10^o}(0)}$=(1.1$\pm$0.18)$\times
$$10^{-5}$ and $Q$=(4.8$\pm$1.5)$\times$$10^{-6}$ (Smoot et al 1992),
thus leading to $b$=(2.3$\pm$0.4).
One can tighten these constraints further by normalising the APM data
to the observed peculiar velocities in the Great Attractor region,
thus explicitly eliminating $b$ (Kashlinsky 1992c). We do this using
the peculiar velocity data at $r$=$40h^{-1}$Mpc from Bertschinger et al (1990)
The numbers
for the quadrupole anisotropy, $Q$, and the smoothed MBR correlation amplitude,
$\sqrt{C_{10^o}(0)}$, we obtain are consistent within the error bars
with the COBE
results and can then be interpreted as supporting the standard inflationary
picture.

We emphasize at the same time that inflation would be inconsistent
with the COBE results and the large-scale structure data if either the
transition to the HZ regime is less sharp than assumed in (2) or if more power
is found on scales that currently cannot be probed by galaxy samples.
Furthermore, the presence of the gravitational
wave background, which is an inevitable consequence of inflation as discussed
by Paul Steinhardt in these proceedings, would produce an extra contribution
to (6) and lead to (much) \underline{larger}
values of $b$ required by matching the APM data to COBE.
\section{High-$z$ objects: constraints on $P(k)$ on small scales}
On scales which are non-linear \underline{today},
$r<8h^{-1}$Mpc, the APM data and eq.(2)
give little direct information on the \underline{primordial} form of $P(k)$.
However, inflation makes also very robust predictions what this form should
be once $P(k)$ is normalised to the large-scale data.
Precisely because inflationary models have no free parameters
($b$ can now be fixed as discussed above), the early
evolution of density fluctuations would lead to
a unique (for a given $\Omega_{CDM}$, $\Omega_{HDM}$ and $\lambda$) transfer
function thus also constraining the small scale power spectrum
which is responsible for collapse of objects at high $z$. This was discussed by
Cavaliere and Szalay (1986) and Efstathiou and Rees (1978) and in the context
of the inflationary models normalised to the large-scale data by Kashlinsky
(1993). We briefly review the results here.

As discussed above in order to account for all the data,
inflationary models have to
be normalised to the power spectrum seen in the APM catalog. I.e. the zero
crossing of the two-point correlation function
should occur on scale $ r $$ \simeq $$
2.5k_0^{-1} $$ \simeq $100$ - $$ 150h^{-1} $Mpc instead of 30$(\Omega h)^{-1}
h^{-1}$Mpc for the standard, $\Omega$=1 and $h$=$\frac{1}{2}$, CDM model.
Two ways have been suggested to overcome this problem and to increase
the power on large scales: 1) Introduce the
cosmological constant $\Lambda$$\equiv$$3H_0^2\lambda$ such that
$\Omega$+$\lambda$=1 ; in Efstathiou et al (1988)
it is shown that such model with
$\Omega h$$\sim$0.1$-$0.2 would produce the large scale power seen in the APM.
2) Introduce two types of dark matter: HDM+CDM; e.g. Davis et al (1992) and
Taylor and Rowan-Robinson (1992)
shows that if $\Omega_{HDM}$$\simeq$0.3 with the remaining contribution
to $\Omega_{total}$=1 coming from CDM, such model gives good fits to a variety
of large-scale structure data.

However, CDM/inflationary models would at the same time suppress the small
scale
power and hence have difficulty
in accounting for the observed objects at $z$$>$3$-$4. Indeed, for the
$\Omega$$+$$\lambda$=1 models the transfer function is given by
$T(k)$$=\{1+[ak+(bk)^{3/2}+(ck)^2]^\nu\}^{-1/\nu}$, where $\nu$=1.13 and
$a$=$6.4 (\Omega h)^{-1} h^{-1}$Mpc; $b$=$3.2 (\Omega h)^{-1} h^{-1}$Mpc;
and $c$=$1.7 (\Omega h)^{-1} h^{-1} $Mpc (Bond and Efstathiou 1984).
The range of $T(k)$$\simeq$1 or the effective power
spectrum index $n$=1 corresponds to scales where the Harrison-Zeldovich
form of the power spectrum got preserved and so $P(k)$ enters the
Harrison-Zeldovich regime
for scales $>$$2a$. On smaller scales the power index
varies from $n$$\simeq$1 through $n$$\simeq$$-1$, required by the APM data,
to $n$$\simeq$$-3$ for scales $\ll$$c$. The scales where
$n$$\simeq$$-3$ correspond to very little small scale power and this
suppresses collapse of fluctuations (and galaxy formation) until a fairly
low $z$ for CDM models.
Lowering $\Omega h$ increases $a$ and
thus can provide the power found in the APM survey; at the same time this
increases $c$$\propto$$(\Omega h)^{-1}$ and further suppresses early
collapse of density fluctuations on the relevant scales.
A similar effect would be
achieved if part of the contribution to $\Omega_{total}$ is due to HDM
(van Dalen and Schaeffer 1992).

The predictions of so normalised CDM/inflationary models can and must be
compared to the
observational data on 1) QSOs at $z$$\geq$$4.5$; 2) the recently
found protogalaxies at $z$$\sim$4; and 3) the protocluster-size object
recently discovered by Uson et al (UBC) (1991). As discussed (Kashlinsky
1993) one can avoid the difficulty with the currently
observed QSO abundances ($z\simeq 4.5$) mainly
because of the freedom one has in determining their total collapsed masses
(cf. Nusser and Silk 1993). But the data on the high-$z$ galaxies ($z\simeq 4$
with the \underline{total} collapsed masses of $>3\times 10^{12}M_\odot$
as the data indicate, see e.g. Chambers and Charlot 1990 and the references
cited
therein) would be difficult to account for on the basis on the modified
CDM models. In other words, it may be
difficult within the framework of CDM models to
account simultaneously for 1) large-scale optical data; 2) COBE results and
3) high-$z$ objects. I.e. if one normalises CDM
models to the large-scale and COBE data, by lowering $\Omega h$ and putting
in $\lambda$=1$-$$\Omega$ or by having HDM as well as CDM, thereby
reducing the small-scale power of the density field, one should then expect
to see 1) significant reduction in the QSO number densities at $z$$>$(4$-$6)
in any modified CDM models;
2) no protogalaxies collapse at $z$$\geq$4; and 3)
no protocluster-size objects, such as the UBC object, at $z$=3.4.
Thus the data on the high-$z$
galaxies may require a power spectrum that has more power on small scales
than CDM models,
e.g. one that has $n$$\sim$$-1$ on scales down to at least
$10^8 M_\odot$, which
scale-invariant inflationary models cannot provide.  The existence of the UBC
object would put even stronger constraints on hierarchical models: indeed,
this object corresponds to the comoving scale $\sim$$8h^{-1}$Mpc;
this fixes its r.m.s. density contrast to be $b^{-1}$ at the present epoch
almost independent of $P(k)$. Its existence at $z$=3.4 would in e.g. CDM
models correspond to a $>$10$-$$\sigma$ fluctuation and one should not see
any of such objects within the horison. To reduce this number
can be achieved by \underline{both} 1) requiring the
light to trace mass, i.e. $b$=1, and 2) making the Universe open, which would
slow down collapse of fluctuations out to and lead to structures forming by
$z_{in}$$\simeq$$\Omega^{-1}$$-$1 as
opposed to $z_{in}$$\simeq$$\Omega^{-\frac{1}{3}}$$-$1 for
$\Omega$$+$$\lambda$=1 models. Such models would require to reconsider the
validity of the standard inflationary assumptions.
\section{Conclusions}
In this review we have discussed the shape of the
\underline{primordial} power spectrum and the values of the cosmological
parameters implied by the data. We have shown that the peculiar velocity
field is in good agreement with that predicted by galaxy correlation data and
that this suggests that the bias factor is constant at least in the linear
regime. The comparison also allows one to estimate $\Omega^{0.6}/b$ and the
data is consistent with $0.2<\Omega^{0.6}/b \leq 0.5$. If the Uinverse turns
out open, it would be impossible to fine-tune inflation as in that case the
super-horison scale inhmogenieties would induce, via the Grischuk-Zeldovich
effect, MBR quadrupole in excess of the COBE data.
We further discussed how 1)the microwave background
data from COBE, 2) the optical data on the distribution of galaxies at $z
\sim0$, and 3) the data on high-$z$ \underline{galaxies} constrain the
primordial $P(k)$ over a range of scales from $<1h^{-1}$Mpc to $>1000h^{-1}$Mpc
and the implications of \underline{all} the data for inflationary scenario(s).
\vspace{1pc}
Bahcall, N. and Soneira, R. 1983, {\it Ap.J.}, {\bf 270}, 20.\\
Collins, C.A. et al 1992, {\it Ap. \ J.}, {\bf 254}, 295.\\
Bertschinger, E. et al 1990, {\it Ap. \ J.}, {\bf 364}, 370.\\
Bond, J.R. and Efstathiou, G. 1984, {\it Ap. \ J.}, {\bf 285}, L45.\\
Cavaliere, A. and Szalay, A. 1986, {\it Ap. \ J.},{\bf 311}, 589.\\
Chambers, K.C. and Charlot, S. 1990, {\it Ap. \ J. \ Lett.}, {\bf 348}, L1.\\
Davis, M. et al 1992, {\it Nature}, {\bf 359}, 393.\\
Efstathiou, G. and Rees, M. 1988, {\it MNRAS}, {\bf 230}, 5p.\\
Efstathiou, G. et al 1990, {\it Nature}, {\bf 348}, 705.\\
Ellis, G. 1988, {\it Class. \ Quantum \ Grav.}, {\bf 5}, 891.\\
Fisher, K. et al 1993, {\it Ap. \ J.}, {\bf 402}, 42.\\
Frieman, J., Kashlinsky, A. and Tkachev, I. 1993, {\it in preparation}.\\
Gorski, K. 1991, {\it Ap. \ J. \ Lett.}, {\bf 370}, L5.\\
Grischuk, L. and Zeldovich, Ya.B. 1978, {\it Sov. Astron.}, {\bf 22}, 125.\\
Juszkiewicz, R. et al 1987, {\it Ap. \ J. \ Lett.}, {\bf 323}, L1.\\
Kaiser, N. 1984, {\it Ap. \ J. \ Lett.}, {\bf 284}, L9.\\
Kaiser, N. 1988, {\it MNRAS}, {\bf 231}, 149.\\
Kashlinsky, A. 1987, {\it Ap.J.}, {\bf317}, 19.\\
Kashlinsky, A. 1991a, {\it Ap. \ J. \ Lett.}, {\bf376}, L5.\\
Kashlinsky, A. 1991b, {\it Ap. \ J. \ Lett.}, {\bf 383}, L1.\\
Kashlinsky, A. 1992a, {\it Ap. \ J. \ Lett.}, {\bf 386}, L37.\\
Kashlinsky, A. 1992b, {\it Ap. \ J. \ Lett.}, {\bf 387}, L1.\\
Kashlinsky, A. 1992c, {\it Ap. \ J. \ Lett.}, {\bf 399}, L1.\\
Kashlinsky, A. 1993, {\it Ap. \ J. \ Lett.}, {\bf 406}, L1.\\
Kellerman, K. 1993, {\it Nature}, {\bf 361}, 134.\\
Kellerman, K. 1993, {\it these proceedings}.\\
Lee, Y.-W. 1992, {\it Astron. J}, {\bf 104}, 1780.\\
Lee, Y.-W. 1993, {\it these proceedings}.\\
Maddox, S. et al 1990, {\it MNRAS}, {\bf 242}, 43p.\\
Nusser, A. and Silk, J. 1993, {\it Ap. \ J. \ Lett.}, {\bf 411}, L1.\\
Peacock, J. 1991, {\it MNRAS}, {\bf 253}, 1p.\\
Peebles, P.J.E. 1987, {\it Nature}, {\bf 327}, 210.\\
Peebles, P.J.E. 1989, {\it Ap. \ J. \ Lett.}, {\bf 339}, L5.\\
Picard, A. 1991, {\it Ap. \ J. \ Lett.}, {\bf 368}, L7.\\
Postman, M. et al 1986, {\it Astron. J.}, {\bf 91}, 1267.\\
Saunders, W. et al 1991, {\it Nature}, {\bf 349},32.\\
Smoot, G. et al 1992, {\it Ap. \ J. \ Lett.}, {\bf 396}, L1.\\
Sutherland, W. 1988, {\it MNRAS}, {\bf 234}, 159.\\
Steinhardt, P. 1990, {\it Nature}, {\bf 345}, 47.\\
Steinhardt, P. 1993, {\it these proceedings}.\\
Szalay, A. et al 1989, {\it Ap. \ J. \ Lett.}, {\bf 339}, L5.\\
Taylor, A.N. and Rowan-Robinson, M. {\it Nature}, {\bf 359}, 396.\\
Turner, M. 1991, {\it Phys. \ Rev. \ D}, {\bf 44}, 3737.\\
Tully, B. 1993, {\it these proceedings}.\\
Uson, J. et al 1992, {\it Phys. \ Rev. \ Lett.}, {\bf 67}, 3328.\\
van Dalen, A. and Schaeffer, R. 1992, {\it Ap. \ J.}, {\bf 398}, 33.\\
Vogeley, M. et al 1992, {\it Ap. \ J. \ Lett.}, {\bf 391}, L5.\\
Yahil, A. 1993, {\it these proceedings}.
\end{document}